\documentclass[twocolumn,superscriptaddress,amsmath,amssymb,showpacs,pre]{revtex4-1}
\usepackage{graphicx}
\usepackage{color}
\renewcommand{\phi}{\varphi}

\begin{document}

\title{Effects of dopants on the glass forming ability in Al-based metallic alloy}

\author{Yang Sun}
	\affiliation{Ames Laboratory, US Department of Energy, Ames, Iowa 50011, USA}
\author{Feng Zhang}
	\affiliation{Ames Laboratory, US Department of Energy, Ames, Iowa 50011, USA}
\author{Lin Yang}
	\affiliation{Ames Laboratory, US Department of Energy, Ames, Iowa 50011, USA}
	\affiliation{Department of Physics, Iowa State University, Ames, Iowa 50011, USA}
\author{Huajing Song}
	\affiliation{Ames Laboratory, US Department of Energy, Ames, Iowa 50011, USA}
\author{Mikhail I. Mendelev}
	\affiliation{Ames Laboratory, US Department of Energy, Ames, Iowa 50011, USA}
\author{Cai-Zhuang Wang}
	\affiliation{Ames Laboratory, US Department of Energy, Ames, Iowa 50011, USA}
	\affiliation{Department of Physics, Iowa State University, Ames, Iowa 50011, USA}
\author{Kai-Ming Ho}
	\affiliation{Ames Laboratory, US Department of Energy, Ames, Iowa 50011, USA}
	\affiliation{Department of Physics, Iowa State University, Ames, Iowa 50011, USA}

\date{Nov. 5, 2018}

\begin{abstract}
The effect of dopants on the metallic glass forming ability is usually considered based on analysis of changes in the liquid structure or thermodynamics. What is missing in such considerations is an analysis of how a dopant changes the properties of the crystal phases which can form instead of the glass. In order to illuminate this aspect we performed molecular dynamics simulations to study the effects of Mg and Sm dopants on the crystal nucleation in Al. The simulation data were found to be consistent with the experimental observations that addition of Mg to Al does not lead to vitrification but addition of only 8\% Sm does. The significant effect of Sm doping was related to the intolerance of Al to this dopant. This leads to increase in the solid-liquid interfacial free energy, and therefore, to increase in the nucleation barrier and to dramatic decrease in the nucleation rate. The intolerance mechanism also significantly affects the growth kinetics.

\end{abstract}

\maketitle

\section{Introduction}
The glass forming ability (GFA) of metallic alloys has been discussed for several decades \cite{Inoue2000,Cheng2011}. One of the main goal here is to \textit{a priori} predict a combination of dopants which will allow one to make the glass more stable against a thermal treatment \cite{Inoue1993,Schroers2001,Egami1997,Laws2015,Suryanarayana2009,Russo2018}. Several empirical rules ranging from over simplistic approaches based on atomic radii ratios or liquid solution formation enthalpy criteria to very sophisticated analysis of the liquid structure were proposed. What seems to be missing in such considerations is an analysis of how a dopant changes the properties of the crystal phases which can form instead of the glass. However, this may be the key to the GFA problem because a glass can be formed and be stable only when a crystal counterpart cannot nucleate or grow \cite{Turnbull1969}. Major advances were achieved in the last two decades in the classical nucleation theory (CNT) by combining it with the molecular dynamics (MD) simulation \cite{Sosso2016}. However, the glass forming alloys represent a special challenge: naturally these are the systems where the nucleation is never observed during the limited time (less than tens of microseconds) of MD simulations. Recently we proposed a persistent-embryo method (PEM) to overcome this time limitation \cite{Sun2018a}. In the present study, we employ this method to compare the effects of Mg and Sm dopants on the nucleation rate in Al and show that the PEM allows to predict that addition of Mg does not lead to any considerable change in the GFA while addition of Sm does. On contrary, we find that none of the conventional analysis of the liquid structure and viscosity can explain the significant effect of Sm dopant on the GFA. 

	Al-rare earth (Al-RE) alloys are a typical example of binary alloys where a rather small addition of a dopant can dramatically change the glass forming ability and the crystallization behavior upon cooling \cite{Shen2017}. While pure Al cannot be obtained in the form of glass, addition of only 8 at.\% Sm doping leads to a possibility of producing a glass \cite{Inoue1988}. Several explanations of this phenomenon have been proposed: the mixing enthalpy \cite{Inoue1998}, packing efficiency \cite{Miracle2003,Miracle2004,Yang2010}, the icosahedral ordering in the liquid \cite{Bokas2016}, etc. While these studies brought some insight into the formation of Al-based metallic glasses \cite{Shen2017} we will show below that the combination of the CNT and MD simulation provides a much more straightforward approach: we can simply compare how small additions of Sm and Mg change the nucleation and growth of the face-centered cubic (fcc) phase in the undercooled Al liquid. The reason for choosing Mg is associated with the fact that Mg has a similar atomic radius (160 pm) as the RE elements (164-180 pm \cite{Miracle2006}) and similar coordination number (~16) in Al-rich liquid. However, in contrast to the marginal GFA of the Al-RE alloys, no glass formation has been reported for the Al-rich Al-Mg alloys. Therefore, the different effects of Mg and Sm dopants provide a perfect testbed to reveal the mechanism of the vitrification. 
	
	The rest of the paper is organized as follows. First, we consider the effect of Sm and Mg on the liquid structure and viscosity and show that there is no striking difference between the effects of these two dopants. Then, we consider their effect on the nucleation ability and show that the effect of Sm is much larger. In order to explain this observation, we turn to the analysis of the effect of these dopants on bulk driving force and the solid-liquid interface (SLI) properties. We will show that while there is no much difference in the effect of these dopants on the bulk driving force, the presence of Sm in the liquid increases the SLI free energy because the solid Al is intolerant to this dopant. Finally, we will discuss all of these findings in relation to the GFA.
\vfill
\section{Effects of dopants on the bulk liquid properties}
\begin{figure}
\includegraphics[width=0.5\textwidth]{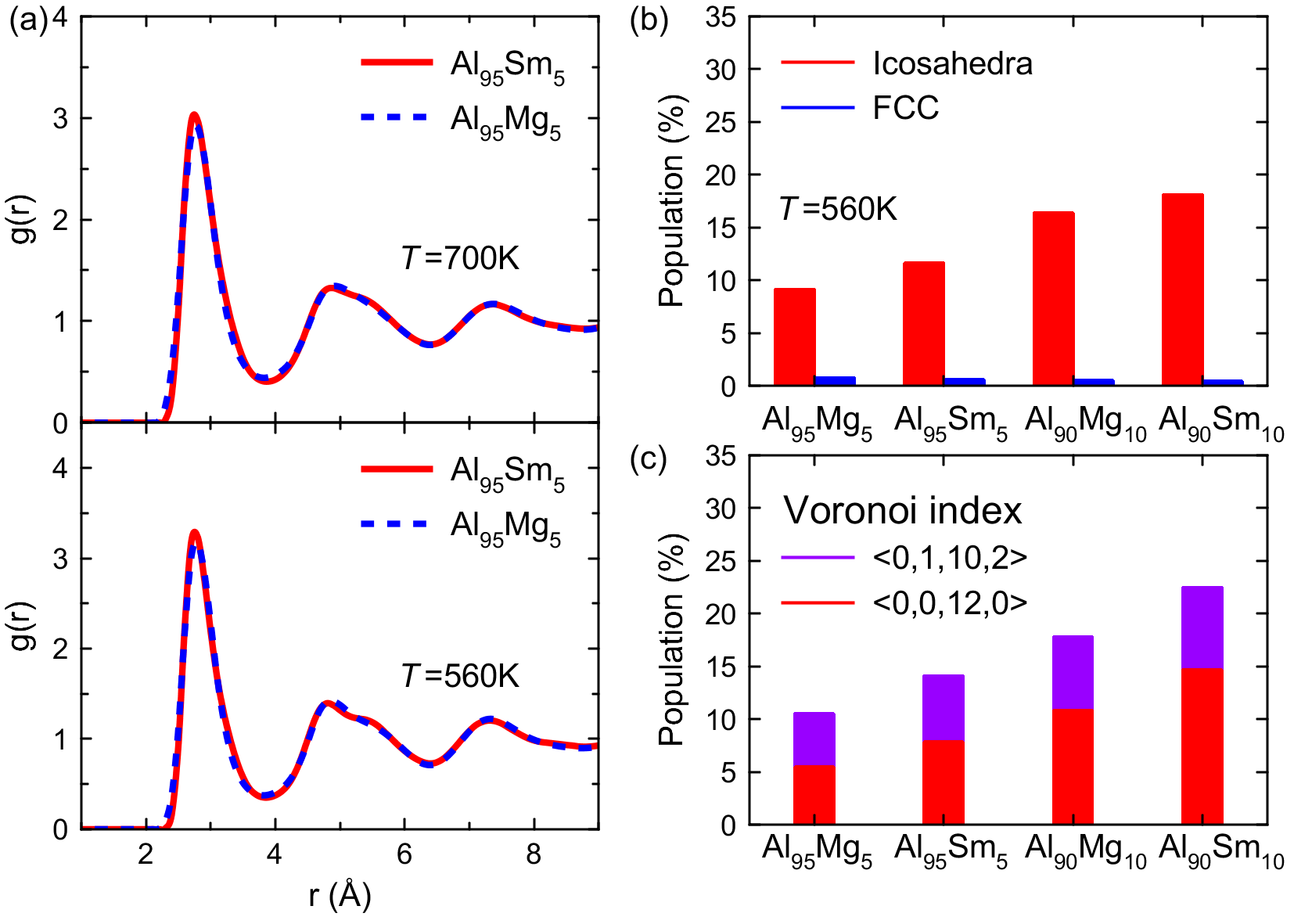}
\caption{\label{fig:fig1} The comparison of liquid structures of the Al-Sm and Al-Mg alloys.(a) The total pair correlation functions.(b) The icosahedra and FCC SRO of Al-centered clusters in the Al-Sm and Al-Mg alloys with the dopant concentrations of 5 and 10 at.\% at 560 K obtained by the cluster alignment. The score cutoff to identify the SRO was set 0.16 \cite{Sun2016}. (c) The population of the top two Voronoi polyhedra $\langle$0,0,12,0$\rangle$ and $\langle$0,1,10,2$\rangle$ in the Al-Sm and Al-Mg alloys. Both two Voronoi polyhedra are associated with the icosahedral ordering \cite{Cheng2011}.}
\end{figure}

To understand what causes the difference in the glass-forming ability of Al-Mg and Al-Sm alloys, we first investigate how the Mg and Sm dopants change the liquid structure. All MD simulations in the present study were performed using the GPU-accelerated LAMMPS code \cite{Brown2011,Brown2012,Brown2013}. The interatomic interaction was modelled using the Finnis-Sinclair potentials  \cite{Finnis1984} developed for the Al-Sm alloys in Ref. \cite{Mendelev2015} and for the Al-Mg alloys in Ref. \cite{Mendelev2009}. In Fig. ~\ref{fig:fig1}(a), the pair correlation functions (PCF) are shown for the 32,000-atom liquid models of Al$_{\text{95}}$Sm$_{\text{5}}$ and Al$_{\text{95}}$Mg$_{\text{5}}$ alloys at moderate (700 K) and deep undercoolings (560 K). One can see that the PCFs of the two systems are almost identical. We also examined other compositions up to 10 at.\% of the dopant concentration and found the results are almost same as shown in Fig. ~\ref{fig:fig1}(a). To further understand the liquid structure we analyzed the short-range order (SRO) in the Al-Sm and Al-Mg alloys. We also included models of Al$_{\text{90}}$Sm$_{\text{10}}$ and Al$_{\text{90}}$Mg$_{\text{10}}$, as the Al$_{\text{90}}$Sm$_{\text{10}}$ metallic glass can be formed experimentally \cite{Inoue1998}. Two widely used analysis are performed to identify the SRO in these samples at T=560 K: the cluster-template alignment method \cite{Fang2010} and the Voronoi tessellation analysis \cite{Sheng2006}. The cluster-template alignment method aligns the atomic clusters to the perfect templates to obtain the alignment score, which describes the minimal root-mean-square deviations between the clusters and the template. Here, the popular icosahedral SRO (ISRO) and the main competing FCC SRO are considered. The score cutoff to identify the SRO \cite{Sun2016} was 0.16. The populations of icosahedral and fcc clusters in the Al-Sm and Al-Mg alloys are shown in Fig. ~\ref{fig:fig1}(b). The Voronoi tessellation analysis characterizes the local atomic environment with the Voronoi index $\langle$n3, n4, n5, n6$\rangle$, where ni denotes the number of i-edge faces of the Voronoi polyhedron (VP). The analysis shows that $\langle$0,0,12,0$\rangle$ and $\langle$0,1,10,2$\rangle$ are always the top two types of the VP for the Al-centered clusters in both the Al-Sm and Al-Mg systems. The VP with index $\langle$0,0,12,0$\rangle$ is icosahedron and $\langle$0,1,10,2$\rangle$ is considered as the distorted icosahedron \cite{Cheng2011}. Therefore, the two types of VP are both associated with ISRO. Comparing the populations in Fig. ~\ref{fig:fig1}(b) and (c), we see that both methods lead to the same conclusion that minor Sm- and Mg-doped Al alloys have a similar icosahedral ordering at deep undercoolings (although at the same concentrations, addition of Sm leads to slightly higher ISRO). The FCC-type clusters are very rare and their fraction is almost independent on the minor Sm or Mg doping concentrations in the liquid Al.

\begin{figure}
\includegraphics[width=0.5\textwidth]{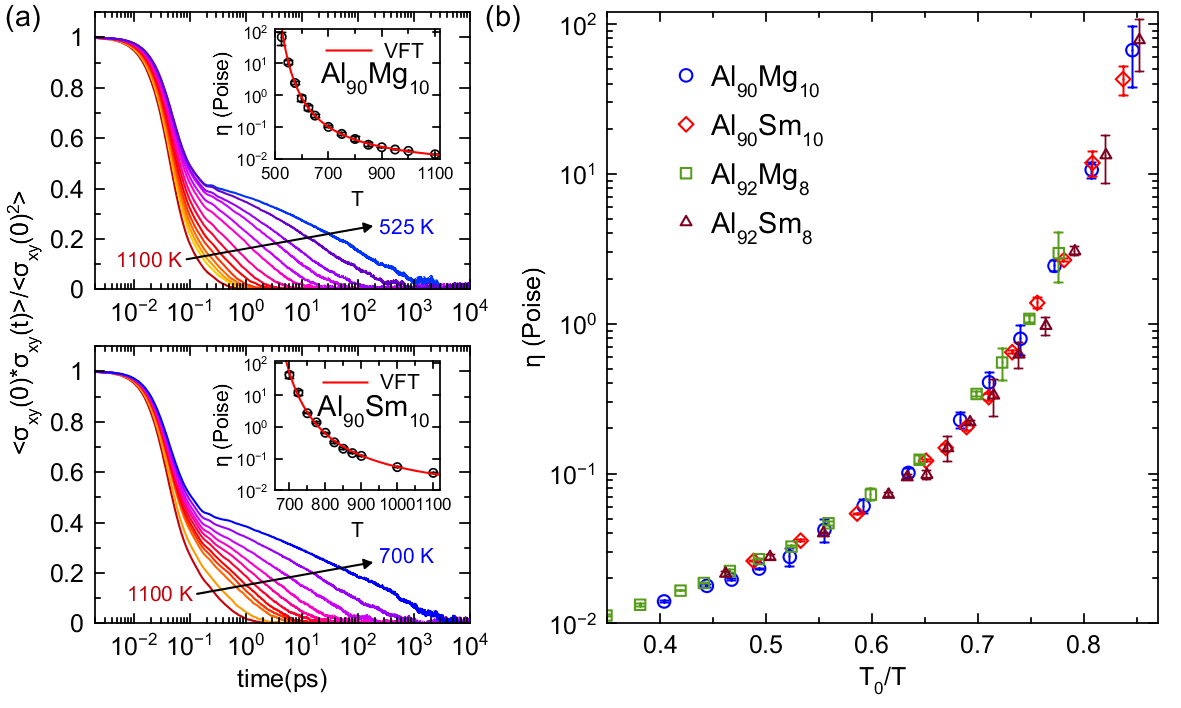}
\caption{\label{fig:fig2} The comparison of bulk liquid kinetics of Al-Sm and Al-Mg alloys.(a) Normalized stress autocorrelation functions of Al$_{\text{90}}$Mg$_{\text{10}}$ (upper) and Al$_{\text{90}}$Sm$_{\text{10}}$ (lower) melts at various temperatures. The colors from red to blue indicate the decreasing temperatures. The insert shows the computed shear viscosity using the Green-Kubo relation and the temperature dependences are fitted using the VFT equation. The fitting functions are $\eta=0.0038\exp(\frac{841.0}{T-444.2})$ and $\eta=0.0042\exp(\frac{1060.6}{T-586.9})$ for the Al$_{\text{90}}$Mg$_{\text{10}}$ and Al$_{\text{90}}$Sm$_{\text{10}}$ alloys, respectively. The error bar was obtained by averaging over all the autocorrelation functions of three off-diagonal components of the stress tensor. (b) The viscosity as function of $T_0$-scaled temperature for the Al-Mg and Al-Sm alloys. Note that we follow the common practice by setting the diverging temperature $T_0$ equal to the glass transition temperature since the two are close for many metallic liquids \cite{Mauro2014,Angell1995,Bohmer1992}.}
\end{figure}

We now turn to the investigation of the kinetic properties of the bulk liquid focusing on the liquid fragility, which measures how quickly the viscosity of a liquid increases upon cooling. Recent progress in the fragility studies have brought a very valuable insight into its relation with the liquid structure and thermodynamics properties \cite{Martinez2001,Mauro2014,Jaiswal2016}. On the other hand, the correlation between the fragility and the glass-forming ability is still under debate \cite{Mauro2014,Fan2004} . Here, the shear viscosities were computed in the NVT MD simulations via the autocorrelation functions of the stress tensor using the Green-Kubo relation \cite{Allen2017} 
\begin{equation}
\eta = \frac{V}{k_{B}T} \int_{0}^{dt} dt \langle \sigma_{xy}(0)\sigma_{xy}(t) \rangle
\label{eos},
\end{equation}
where $\sigma_{xy}$ is the off-diagonal x-y component of the stress tensor, V is the volume of the liquid. The normalized shear autocorrelation functions of both Al$_{\text{90}}$Mg$_{\text{10}}$ and Al$_{\text{90}}$Sm$_{\text{10}}$ are shown in Fig.~\ref{fig:fig2}(a). Both systems show very similar trends for the shear relaxation with the temperature: the shear relaxation is almost exponential at higher $T$, while it becomes highly non-exponential when the system is cooled down to lower $T$ regime. The decrease in temperature leads to rapid increase in the correlation time and develops a “bump” at the correlation time range between 0.2 and 0.3 $ps$, which can be attributed to the effect of the boson peak vibrations in the glassy and deep undercooled liquid states \cite{Han2011,Jaiswal2015}. Figure~\ref{fig:fig2}(a) demonstrates that the obtained temperature dependences of the viscosities can be well fitted to the Volger-Fulcher-Tafmmann (VFT) equation $\eta = \eta_{0} \exp(\frac{B}{T-T_0})$. Since our data were mostly obtained at high temperature, we employed the kinetic strength, which is defined as $D^*=\frac{B}{T_0}$ , to quantify the fragility \cite{Mauro2014}. We found that the fragilities of Al$_{\text{90}}$Mg$_{\text{10}}$ and Al$_{\text{90}}$Sm$_{\text{10}}$ are very close to each other ($D^*=1.89$ and $D^*=1.81$, respectively). Figure~\ref{fig:fig2}(b) shows the viscosities as a function of the $T_0$-scaled temperature including lower doping concentration 8\%. The examination of this figure clearly demonstrates that both systems exhibit a similar behavior of the kinetics slowdown when approaching the glass transition temperature.

The results of the structures and fragility studies are consistent with each other: the effects of Mg and Sm dopants on the liquid properties are very similar. Therefore, the empirical rules based solely on the analysis of the liquid properties \textit{cannot} explain the difference in the glass formality of the Al-Mg and Al-Sm alloys.

\section{Effects of dopants on the nucleation rate}
Since the dopants only shows very minor effect on the liquid properties, we now employ the PEM simulations to study the nucleation rate for the understanding of the glass formation. The PEM utilizes the main CNT \cite{Sun2018a} concept that the homogeneous nucleation happens via the formation of the critical nucleus in the undercooled liquid. The excess free energy to form a nucleus with N atoms can be presented as 
\begin{equation}
\Delta G = N \Delta \mu + A \gamma
\label{eos},
\end{equation}
where $\Delta\mu(<0)$ is the chemical potential difference between the bulk solid and liquid, $\gamma$ is the solid-liquid interfacial (SLI) free energy density, and $A$ is the SLI area. The competition between the bulk and interface terms leads to a nucleation barrier, $\Delta G^{*}$, when the nucleus reaches the critical size, $N^{*}$. While the CNT usually relies on the assumption that the nucleus has a spherical shape to derive the relation between $\Delta G^{*}$ and $N^{*}$ there is no real need to make this assumption. Instead we present the SLI area as $ A = s ( \frac{N}{\rho_c} ) ^ \frac{2}{3} $, where $\rho_c$ is the crystal density and $s$ is a shape factor. If we make the assumption that $s$ is independent of the nucleus size $N$ (which is weaker than the assumption about the spherical nucleus shape) it is easy to show that  the nucleation barrier, $\Delta G^{*}$, can be written as \cite{Sun2018a}:
\begin{equation}
\Delta G^{*} = \frac{1}{2} \vert \Delta \mu \vert N^{*}
\label{eos}.
\end{equation}
Following Auer and Frenkel \cite{Auer2004}, the nucleation rate can be calculated as
\begin{equation}
J=\rho_L f^{+} \sqrt{\frac{\vert \Delta \mu \vert}{6 \pi k_{B} T N^{*}}} \exp(-\frac{\Delta G^{*}}{k_{B} T})
\label{eos},
\end{equation} 
where $f^+$ is the atomic attachment rate, $k_BT$ is the thermal factor and $\rho_L$ is the liquid density. According to Eq. (4), four quantities ($\rho_L$, $N^\ast$, $\Delta \mu$, and $f^+$) are needed to obtain from the MD simulations to calculate the nucleation rate at a given temperature. The determination of the liquid density, $\rho_L$, is trivial. The chemical potential difference, $\Delta \mu$, can be calculated using the thermodynamic integration based on an alchemical path linking the doped and pure liquids \cite{Yang2018}(see Supplementary Materials for details). The method to determine the critical nucleus size $N^\ast$ was described in detail in Ref. \cite{Sun2018a}. Finally, the attachment rate $f^+$ can be measured in the MD simulation following the diffusion approach proposed by Auer and Frenkel \cite{Auer2004,Auer2001,Espinosa2015}.

\begin{figure}[t]
\includegraphics[width=0.9\textwidth]{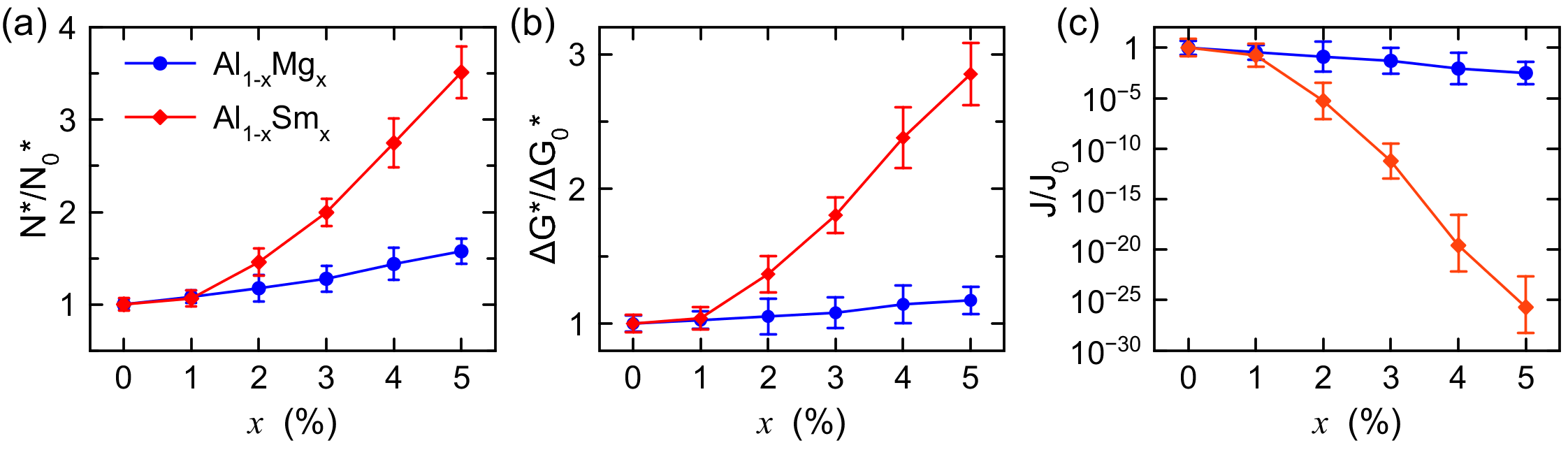}
\caption{\label{fig:fig3} FCC nucleation in the Al-Mg and Al-Sm alloys obtained from the PEM at moderate undercooling $T=700 K$. (a) The critical nucleus size, (b) the nucleation barrier and (c) the nucleation rate as functions of the doping concentrations. All the quantities are scaled by the corresponding values obtained for the pure Al (denoted by subscript “0”).}
\end{figure}

Figure~\ref{fig:fig3} summarizes the nucleation rate data for the Al-Sm alloys obtained in Ref.  \cite{Sun2018b} and the data for the Al-Mg obtained within the present study. At very small dopant concentrations (up to 1 at.\%) the effects of both dopants are about the same. However, at higher concentrations, while addition of Mg leads to only small increase in the critical nucleus size, the addition of Sm changes it much more significantly (3.5 times at 5\% of Sm). This is obviously associated with the fact that the addition of Sm considerably changes the nucleation barrier (see Fig. ~\ref{fig:fig3}(b)). Since the nucleation rate depends exponentially on the nucleation barrier, the addition of only 5 at.\% Sm decreases the nucleation rate by ~25 orders of magnitude, while addition of 5 at.\% Mg decreases the nucleation rate only by 3 orders of magnitude (see Fig.~\ref{fig:fig3}(c)). These results explain the experimental observations: no glass has been produced in the Al rich Al-Mg alloys but only 8\% Sm doping leads to the vitrification in the Al-Sm alloys \cite{Inoue1998}. 
\onecolumngrid

\clearpage

\twocolumngrid

\section{Effects of dopants on the solid-liquid interface properties}
\begin{figure}
\includegraphics[width=0.4\textwidth]{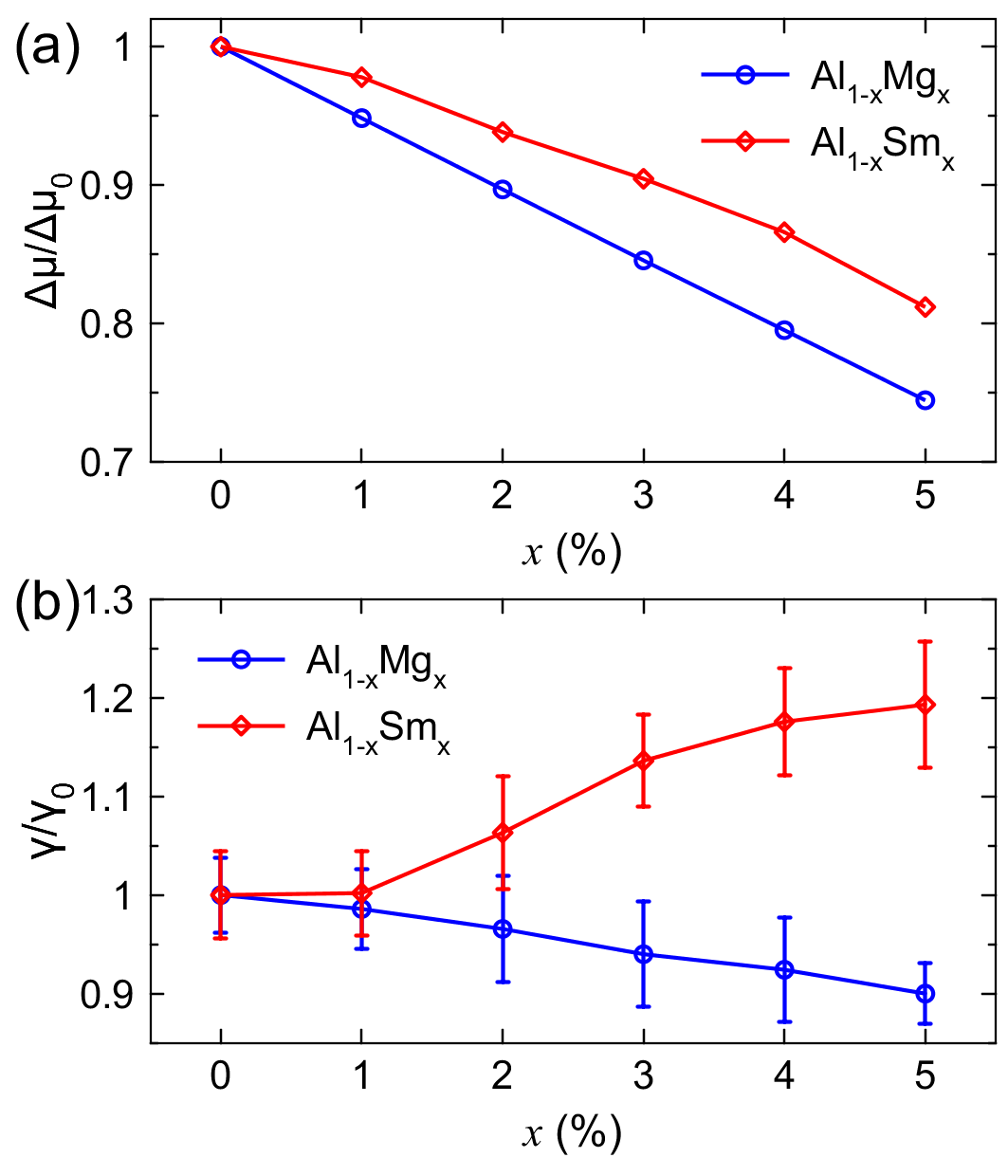}
\caption{\label{fig:fig4} The main quantities controlling the nucleation barrier. (a) The chemical potential difference and (b) the SLI free energy as functions of the doping concentrations at $T=700K$. The quantities are scaled by the corresponding values obtained for the pure Al (denoted by subscript “0”). }
\end{figure}

According to Eqn.(2), the main factors controlling the nucleation barrier are chemical potential difference and the SLI free energy. Figure~\ref{fig:fig4}(a) shows that the Mg dopants decrease the chemical potential difference even more significantly than do the Sm dopants. Since the chemical potential difference provides the positive driving force for the nucleation, the key factors that contribute to the increased nucleation barrier in the Al-Sm alloys must be the increase in the SLI free energy. If we neglect the anisotropy of the SLI free energy, the PEM allows to estimate it as \cite{Sun2018a,Sun2018c}: 
\begin{equation}
\gamma=\frac{3}{2s^\ast}\Delta \mu \rho_{c}^{\frac{2}{3}} N^{*\frac{1}{3}}
\label{eos}.
\end{equation} 
We have determined all quantities in this equation except for the shape factor, $s^\ast$. In order to determine the latter we analyzed the shapes of the critical nuclei obtained from the PEM simulations \cite{Sun2018c}. By constructing the triangulated surface mesh  \cite{Stukowski2014} using the OVITO software package  \cite{Stukowski2010}, the shaper factor, $s^\ast$, was measured according to the surface area and the volume of the nucleus polyhedron. The snapshots of the nucleus were averaged to reduce the effect of the thermal fluctuation on the shape \cite{Sun2018c} (see the Supplemental Material for more details). 

\begin{figure}[h]
\includegraphics[width=0.9\textwidth]{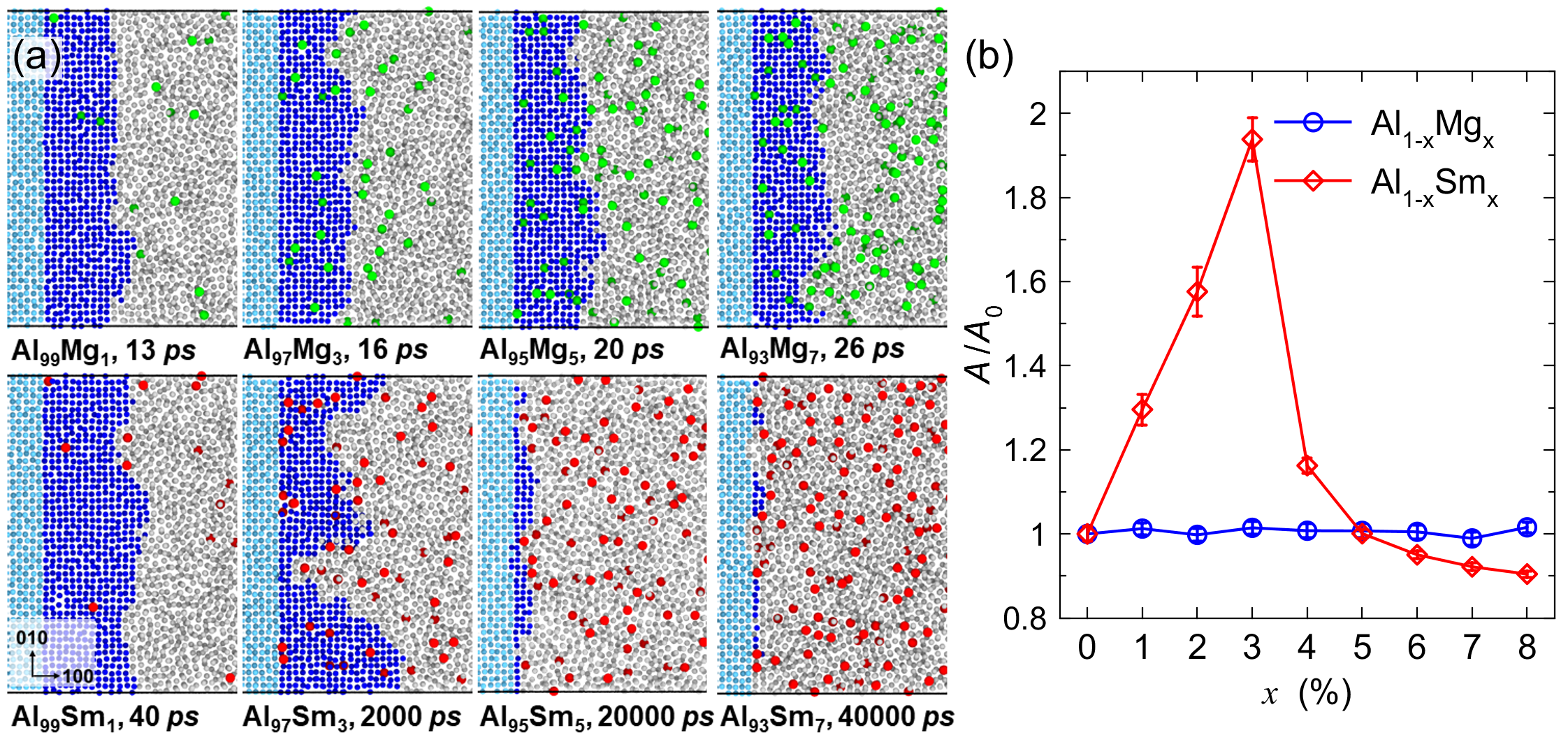}
\caption{\label{fig:fig5} The SLI profiles for AlSm and AlMg. (a) The snapshots at the SLI for the Al-Mg (upper panel) and Al-Sm (lower panel) alloys. Two fcc layers along the in-plane direction (001) are shown for clarity. The grey atoms are liquid-like Al atoms. The blue are the solid-like Al atoms while light blue indicates the initial bulk fcc phase. The red atoms are Sm and green are Mg. For all the Al-Mg, as well as Al$_{\text{99}}$Sm$_{\text{1}}$ and Al$_{\text{97}}$Sm$_{\text{3}}$ alloys, the snapshots were taken at the moment when the as-grown solid phase reaches the similar population (~25,000 atoms). For the Al$_{\text{95}}$Sm$_{\text{5}}$ and Al$_{\text{93}}$Sm$_{\text{7}}$ alloys, the snapshots were taken after significate longer-time simulation. (b) The interfacial area as function of doping concentration. The quantities are scaled with the interfacial area in the pure Al system. }
\end{figure}

We now turn to the question why the Mg and Sm dopants have such a different effect on the SLI free energy. The critical nuclei at $T=700K$ are rather small ($\sim$150-550 atoms) such that it is very difficult to analyze the SLI profiles with such a limited length scale. On contrary, a simulation of a flat interface can provide ample statistics for the SLI curvatures and profiles. Therefore, we performed a series of the MD simulations at $T=700 K$ where initially one part of the simulation cell contained the fcc Al with the [100] direction parallel to the $x$-axis and the other part contained either Al$_{1-x}$Sm$_{x}$ or Al$_{1-x}$Mg$_{x}$ liquid alloy (see the Supplemental Material for simulation details). The MD simulation showed that the Al-Mg liquid (with $x_{\text{Mg}}$ up to 8 at.\%) quickly solidifies into a solid solution with the same Mg concentration as in the liquid phase. As shown in Fig.~\ref{fig:fig5}(a), compared to Mg, the Sm dopants much more effectively slow down the SLI migration even at $x_{\text{Sm}}=1$ at.\% and almost completely stop it (on the MD time scale) at $x_{\text{Sm}}=7$ at.\%. Examination of Fig.~\ref{fig:fig5}(a) shows that the SLI roughness dramatically increases around $x_{\text{Sm}}=3$ at.\% but at the highest Sm concentration we can see that the SLI becomes almost flat. To make a more quantitative description on the interface roughness, we measured the averaged interface area during the SLI migration using the surface mesh method \cite{Stukowski2014}. As shown in Fig. ~\ref{fig:fig5}(b), the interface roughness first quickly increases with addition of Sm reaching the maximum value at 3 at.\% and then quickly decreases reaching the same value as in the pure Al at 5 at.\%. Further addition of Sm leads to slow decrease in SLI roughness.
\onecolumngrid

\twocolumngrid

\begin{figure}
\includegraphics[width=0.5\textwidth]{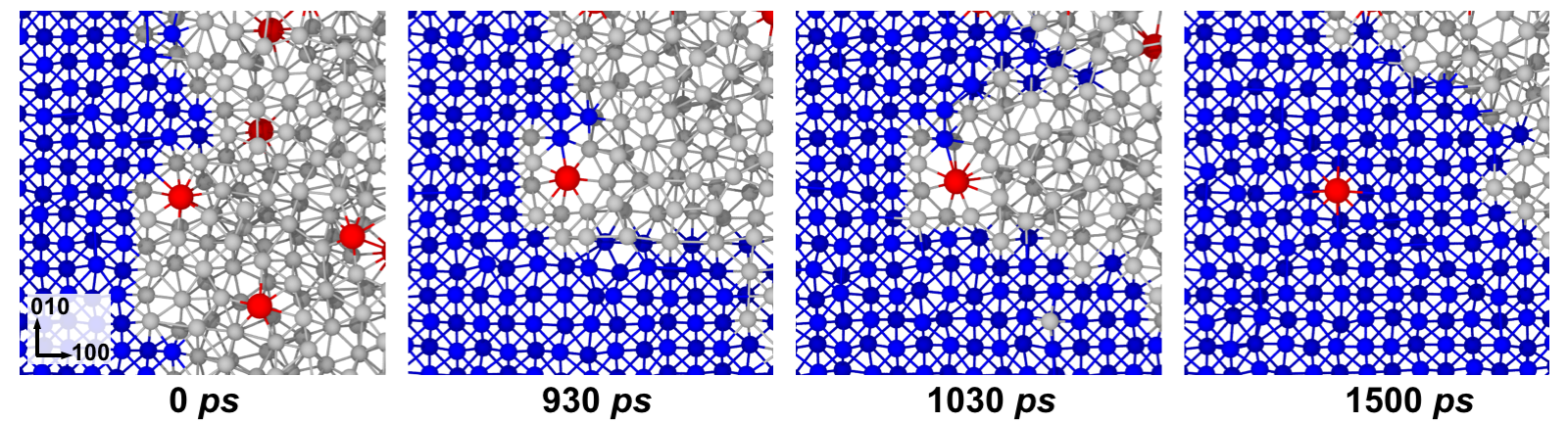}
\caption{\label{fig:fig6} Trapping Sm into the fcc phase growing from the liquid Al$_{\text{97}}$Sm$_{\text{3}}$ alloy. From left to right, it shows a time serial of the Sm atom incorporating into the fcc-Al lattice. }
\end{figure}

While both Mg and Sm dopants seem to be able to be incorporated into the growing fcc-Al phase, they clearly show a drastic different effect on the growth kinetics. In order to get more insight, we carefully investigated the mechanism of the dopant transition into the growing fcc phase. We did not find any interesting features in the case of Mg: it simply goes from the liquid into an fcc site. On contrary, Figure~\ref{fig:fig6} shows that the scenario is more complex in the case of Sm. The fcc phase seems hard to tolerate Sm as the Sm does not directly go into the fcc phase like Mg does. Instead, the fcc phase grows where the liquid phase occasionally does not have Sm atoms. This locally increases the SLI curvature and allows the SLI to surround and capture the Sm atom from the liquid (see also the movie in the Supplemental Material). With increasing Sm concentration in the liquid the probability to find a Sm free region decreases leading to decreasing the SLI roughness. Thus, it is the intolerance of the Al fcc phase to Sm which explains the dependence of the SLI roughness on the Sm content (and low SLI velocity). Moreover, the fact that the SLI becomes flat means that its stiffness becomes very large which is consistent with the previous results that the addition of Sm leads to the increase in the SLI free energy.

\section{Discussion}
The current study shows that the key effect of doping that leads to the glass formation in the Al-Sm alloys is the change in the SLI free energy, rather than any change in the bulk liquid properties. This is inherently caused by the intolerance of the Al crystal lattice to Sm. Keeping this in mind, we now turn our attention back to the crystal nucleation from liquid to confirm this scenario. We first checked all the as-formed critical nucleus from PEM simulation at 700 K. We find there is no Sm atom in any critical nucleus, while there are indeed a few Mg atoms in the critical nuclei. However, at $T=700 K$ no nucleation can happen in the course of the conventional (brute-force) MD simulation (this is why the PEM is needed). Therefore, we lower the temperature to $T=560 K$, at which the bulk driving force is much larger, the critical nucleus size is much smaller and the nucleation rate is much higher such that the nucleation may happen in the course of the brute force MD simulation (see Supplemental Material). By checking the spontaneously formed nuclei, we found that the critical nuclei in the Al-Mg alloys may contain Mg atoms, while only pure Al nuclei forms in the Al-Sm alloys. This confirms that the Al fcc phase is tolerant to Mg but not to Sm. Note that the mechanism of capturing of Sm shown in Fig.~\ref{fig:fig6} cannot operate in the case of the nucleation because the nucleus size is smaller than the fluctuation length necessary for this capturing.

In order to numerically characterize the tolerance of a crystal phase to a dopant we propose to use the tolerance enthalpy, $\Delta H_t$. Consider 4 simulation cells: \#1 contains the perfect fcc Al; \#2 is the same as \#1 except that one Al atom is replaced by a dopant; \#3 contains the pure liquid Al; \#4 is the same as \#3 except that one Al atom is replaced by a dopant. Then the tolerance energy can be defined as
\begin{equation}
\Delta H_t=H_2+H_3-H_1-H_4,
\label{eos},
\end{equation} 
where all quantities in the equation are the total enthalpy of the corresponding simulation cells (not normalized by the number of atoms). To obtain these values we used the simulation cells with periodic boundary conditions in all directions containing 2048 (fcc) or 2000 (liquid) atoms. The energies were averaged over 1 ns. We obtained $\Delta H_t=1.3$ eV/atom for Mg and $\Delta H_t=7.0$ eV/atom for Sm. Thus, the fcc Al is indeed much more tolerant to Mg dopants than to Sm dopants. 

\section{Conclusion}
In summary, we employed the PEM to explain why the addition of Sm dramatically enhances the glass forming ability of Al and the addition of Mg does not. Contrary to traditional approaches which consider the liquid alloy structure or kinetics to predict the GFA, the PEM provides a robust way to answer this question. It shows that effect of Sm on the nucleation rate is many orders of magnitude stronger than the effect of Mg which is in agreement with the experimental observations. The large effect of Sm is related to the fact that the fcc Al phase is not tolerant to this dopant. This leads to increase in the SLI free energy, and therefore, to increases in the critical nucleus size and the nucleation barrier and to dramatic decrease in the nucleation rate. By examining detailed SLI migration, a dopant trapping behavior is revealed and also explained by the dopant intolerance for AlSm system. On contrary the fcc Al phase is rather tolerant to Mg such that it can easily occupy a crystal site in the fcc phase growing from the liquid phase. This is why addition of Mg does not lead to a considerable change in the glass forming ability. A tolerance enthalpy is introduced to characterize the tolerance of the crystal phase to a dopant. This quantity can be simply measured by the atomistic simulation. Further studies are needed to relate this quantity to the SLI free energy and nucleation rate with more glass-forming systems. 

\section{Acknowledgments}
Work at Ames Laboratory was supported by the U.S. Department of Energy (DOE), Office of Science, Basic Energy Sciences, Materials Science and Engineering Division, under Contract No. DEAC02-07CH11358, including a grant of computer time at the National Energy Research Supercomputing Center (NERSC) in Berkeley, CA. The Laboratory Directed Research and Development (LDRD) program of Ames Laboratory supported the use of GPU computing.


\pagebreak
\widetext
\begin{center}
\textbf{\large Supplemental Materials for ``Effects of dopants on the glass forming ability in Al-based metallic alloy''} 
\end{center}

\twocolumngrid
\setcounter{equation}{0}
\setcounter{figure}{0}
\setcounter{table}{0}
\makeatletter
\renewcommand{\theequation}{S\arabic{equation}}
\renewcommand{\thefigure}{S\arabic{figure}}
\renewcommand{\bibnumfmt}[1]{[S#1]}
\renewcommand{\citenumfont}[1]{S#1}
\renewcommand{\thesection}{S\arabic{section}}
\setcounter{section}{0}
\section{Chemical potential difference}
The method to determine the chemical potential difference for the fcc-Al nucleation in the undercooled Al$_{1-x}$Sm$_{x}$ liquid alloys (Sm has no solubility in fcc-Al) has been described in detail in Ref. \cite{sYang2018}. Since Mg is soluble in fcc-Al for the composition range studied in this paper, the chemical potential difference is the free energy difference between the liquid and solid solutions. For both solution phases, the free energy was calculated by transforming the pure Al phase into the corresponding solution phase, using the thermodynamic integration along an  “alchemical” path, as outlined in Ref. \cite{sYang2018}. However, in the case of the fcc solid solution, a special treatment is needed to sample different configurations of Mg dopants in the fcc lattice, in order to establish the thermodynamic equilibrium. This was achieved by allowing swaps between Al and Mg atoms with a Monte-Carlo algorithm during MD simulations, as implemented in the LAMMPS package. 

\section{The shape factor of the nucleus}
\begin{figure}
\includegraphics[width=0.45\textwidth]{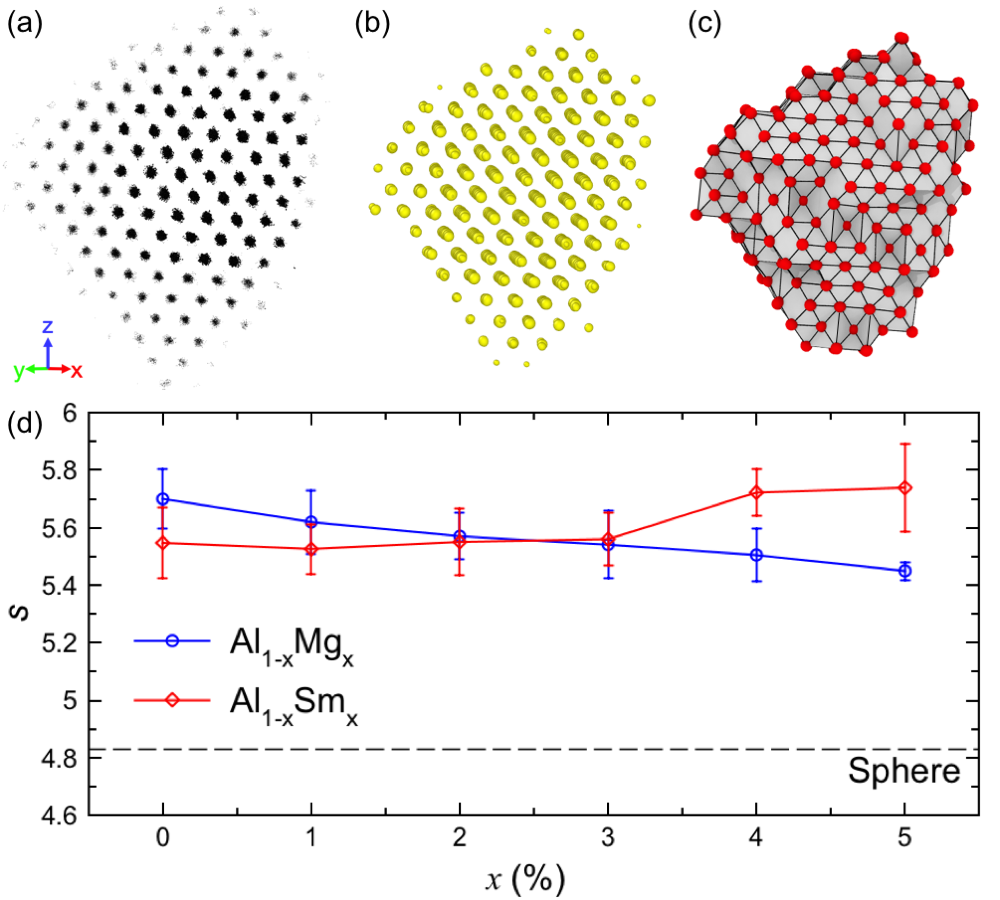}
\caption{\label{fig:figs1} Determination of the nucleus shape. (a) A superposed fcc nucleus configuration from the PEM simulation of Al$_{\text{95}}$Sm$_{\text{5}}$ liquid at $T=700 K$; (b) The atomic-density contour plots corresponding to the distribution in (a); (c) The surface of the nucleus. The connected red points define the surface crystalline sites. The probe sphere radius to construct the surface mesh in the geometric surface reconstruction method \cite{sStukowski2010} is $R_{\alpha}=2.8 \text{Å}$. (d) The measured shape factor as a function of doping concentrations. The dash line indicates the shape factor under spherical shape assumption.  }
\end{figure}

To compute the interfacial free energy with Eq. (5) in the main text, we measured the shape factor s of the critical nucleus obtained from the PEM simulations \cite{sSun2018a,sSun2018b}. To make a statistically sound description of the nucleus shape, we first averaged the nucleus by superposing the configurations collected in a short time interval ($\Delta t_{0}=4 ps$) during the critical plateau (see Ref.  \cite{sSun2018a} for details). As shown in Fig.~\ref{fig:figs1}(a), the superposed configuration shows a clear non-spherical nucleus shape. Since the crystalline order fades at the interfacial region, it results in a weaker atomic distribution at the outer shell of the nucleus. In order to see the averaged nucleus more clearly, a Gaussian smearing scheme \cite{sKusalik1994,sFang2010,sFang2011} was applied to convert the atomic distribution into the atomic density in the 3D space as shown in Fig.~\ref{fig:figs1}(b). By applying a fast-clustering algorithm \cite{sRodriguez2014} on the density profile, we were able to extract the high-density points, which are essentially the as-formed crystalline sites. Then the crystalline sites which were occupied longer than half of the time interval $\Delta t_{0}$ were used to construct the surface of the nucleus by the geometric surface reconstruction method \cite{sStukowski2014} integrated in the OVITO software package \cite{sStukowski2010} as shown in Fig.~\ref{fig:figs1}(c). Finally, the shape factor s was computed based on the surface area $A$ and the volume $V$ of the polyhedron obtained from OVITO as $s = \frac{A}{V^{\frac{2}{3}}}$. In Fig.~\ref{fig:figs1}(d), the shape factors of the critical nucleus are shown as a function of doping concertation in the AlSm and AlMg systems. 

\section{Solid-liquid coexisting simulations}
\begin{figure}
\includegraphics[width=0.45\textwidth]{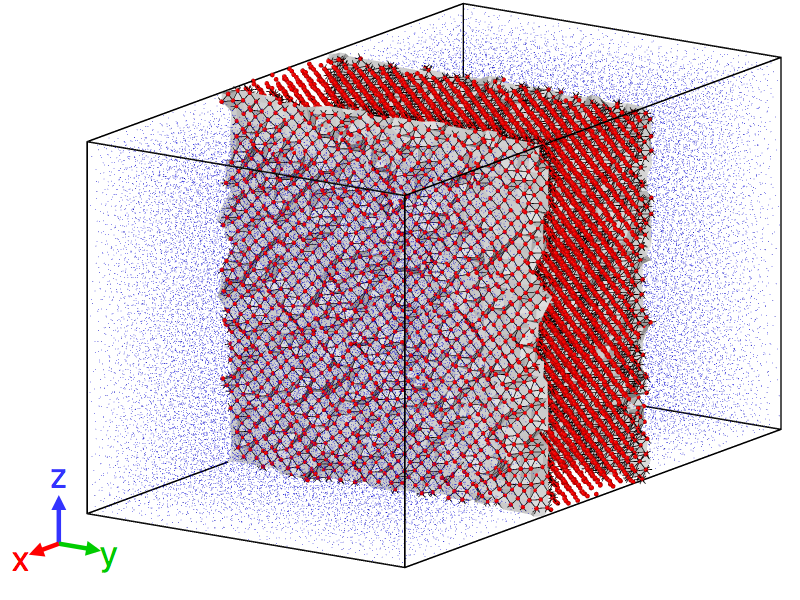}
\caption{\label{fig:figs2} Determination of the interface area in the SLI simulation. The red points are solid fcc atoms, and the blue points are liquid atoms. For clarity, the points for the liquid atoms are shown much smaller than the solid atoms. The solid atoms at the SLI are connected for clarity. The constructed interfaces are shown with grey. The probe sphere radius to construct the surface mesh in the geometric surface reconstruction method \cite{sStukowski2014} is $R_{\alpha}=2.8 \text{Å}$.}
\end{figure}

The initial solid-liquid coexisting model in the simulation was constructed by combining the pure fcc crystal and an undercooled liquid alloy (either Al$_{1-x}$Sm$_{x}$ or Al$_{1-x}$Mg$_{x}$). The periodic boundary conditions were applied in all three directions of the simulation box. All simulations were performed using the Nose–Hoover thermostats. During the simulation, we allow the length of the box in the direction perpendicular to the solid-liquid interface (SLI) to change, resulting in constant area and constant normal pressure: NP$_{x}$T ensemble, with $x$ being the perpendicular direction to the SLI.
The fcc and liquid atoms were distinguished by the cluster alignment method \cite{sFang2010} in which the minimal root-mean-square deviations (alignment score) between the atomic cluster and the perfect fcc template were calculated. The score cutoff to identify crystalline fcc clusters was set as 0.12. With the geometric surface reconstruction method \cite{sStukowski2014} integrated into the OVITO software package \cite{sStukowski2010}, the SLI is constructed according to the configuration of the fcc atoms as shown in Fig.~\ref{fig:figs2}. The SLI area was computed by summing all the facet areas by OVITO software package \cite{sStukowski2010}.

\section{Conventional MD simulation of nucleation at 560K}
\begin{figure}
\includegraphics[width=0.8\textwidth]{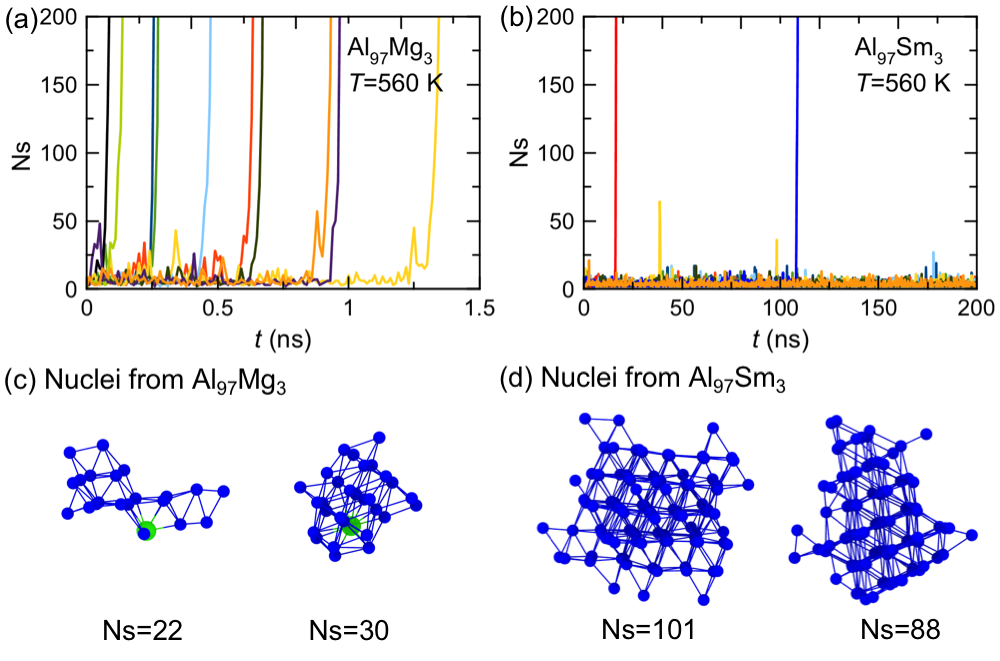}
\caption{\label{fig:figs3} (a) and (b) Numbers of solid atoms as functions of time in 10 independent MD simulations of nucleation at 560 K in the Al$_{\text{97}}$Sm$_{\text{3}}$ and Al$_{\text{97}}$Mg$_{\text{3}}$ alloys, respectively (c) FCC nuclei with embedded Mg atoms obtained from the nucleation simulations of the Al$_{\text{97}}$Mg$_{\text{3}}$ alloy. Only fcc like atoms are shown. Blue atoms are Al and green atoms are Mg. (d) Pure-Al fcc nuclei obtained from two nucleation simulations of the Al$_{\text{97}}$Sm$_{\text{3}}$ alloy. }
\end{figure}

The conventional (brute-force) MD simulations of the nucleation were performed for the Al$_{\text{97}}$Sm$_{\text{3}}$ and Al$_{\text{97}}$Mg$_{\text{3}}$ liquid alloys at the deep undercooling $T=560 K$. The simulation cells contained 32,000 atoms. The isothermal-isobaric (NPT) ensemble and the periodic boundary conditions were applied. The liquid models were initially equilibrated at 1300 K which is well above the Al melting temperature followed by a rapid cooling ($10^{13} K/s$) to the undercooling temperature (560 K). This extremely fast cooling rate prohibits any nucleation during the cooling process. The simulation time was up to 200 $ns$. Due to the stochastic nature of the nucleation process, we independently performed 10 MD runs with different initial configurations for both  Al$_{\text{97}}$Sm$_{\text{3}}$ and Al$_{\text{97}}$Mg$_{\text{3}}$ alloys. As explained in the main text the Sm doping hinders the nucleation of fcc Al much more significantly than the Mg doping. Therefore, it took much longer time to observe the nucleation in the Al$_{\text{97}}$Sm$_{\text{3}}$ alloys than in the Al$_{\text{97}}$Mg$_{\text{3}}$ alloys. As shown in Fig.~\ref{fig:figs3} (a) and (b), all ten Al$_{\text{97}}$Mg$_{\text{3}}$ samples nucleated within 1.5 $ns$, while only two of ten Al$_{\text{97}}$Sm$_{\text{3}}$ samples nucleated within 200 $ns$. With a close inspection of the as-formed nucleus we found Mg atoms incorporated into the fcc nucleus even at the early stage of the nucleation shown in Fig.~\ref{fig:figs3}(c). On contrary, all as-formed nuclei were pure fcc Al in the Al$_{\text{97}}$Sm$_{\text{3}}$ alloy as shown in Fig.~\ref{fig:figs3}(d). 
\onecolumngrid

\clearpage

\twocolumngrid

\end{document}